\documentclass[journal=nalefd,manuscript=letter]{achemso}

\usepackage{chemformula} 
\usepackage[T1]{fontenc} 
\usepackage[separate-uncertainty = true,multi-part-units=single]{siunitx}

\DeclareSIUnit\Molar{M}

\author{Vivian Wang}
\author{Niklas Ermann}
\author{Ulrich F. Keyser}
\email{ufk20@cam.ac.uk}
\affiliation{Cavendish Laboratory, University of Cambridge, 19 JJ Thomson Avenue, Cambridge CB3 0HE, UK}

\title[]
  {Current enhancement in solid-state nanopores depends on three-dimensional DNA structure}


\begin{document}


\begin{abstract}
The translocation of double-stranded DNA through a solid-state nanopore may either decrease or increase the ionic current depending on the ionic concentration of the surrounding solution. Below a certain crossover ionic concentration, the current change inverts from a current blockade to current enhancement. In this paper, we show that the crossover concentration for bundled DNA nanostructures composed of multiple connected DNA double-helices is lower than that of double-stranded DNA. Our measurements suggest that counterion mobility in the vicinity of DNA is reduced depending on the three-dimensional structure of the molecule. We further demonstrate that introducing neutral polymers such as polyethylene glycol into the measurement solution reduces electroosmotic outflow from the nanopore, allowing translocation of large DNA structures at low salt concentrations. Our experiments contribute to an improved understanding of ion transport in confined DNA environments, which is critical for the development of nanopore sensing techniques as well as synthetic membrane channels. Our salt-dependent measurements of model DNA nanostructures will guide the development of computational models of DNA translocation through nanopores.
\end{abstract}


\section{Introduction}
Nanopore sensors are increasingly employed to detect and study a range of single molecules such as DNA, proteins, and viruses \cite{Venkatesan2011}. Identification of a molecule is typically determined by features of the characteristic ionic current signal resulting from electrophoretic translocation of the molecule through a voltage-biased nanopore. The physical and chemical structures of both the nanopore and translocating molecule govern the dynamics of ionic and molecular transport \cite{Keyser2011}. Nanopore-based detection of single DNA molecules has received particular attention for its applications in genomic sequencing and biosensing as well as fundamental biophysical studies \cite{Clarke2009, Li2003}. Interestingly, DNA carries negative charge and due to its counterion cloud may either increase or decrease the ionic current as it translocates across a nanopore (Figure \ref{fgr:setup}). The ionic concentration of the electrolytic solution in which the measurement is performed determines the details of the current change. The concentration at which the transition from current blockage to current enhancement occurs is termed the crossover concentration and was first reported for double-stranded DNA (dsDNA) translocation in inorganic nanotubes \cite{Fan2005}. This phenomenon has since been observed for dsDNA translocation in solid-state nanopores \cite{Smeets2006,Steinbock2012a, Ho2005, Schoch2005, Stein2004, Karnik2005, Siwy2005}. Molecular simulations suggest that ionic current changes in nanopores arise from a number of competing effects including volume exclusion, counterion condensation, and molecular friction between the DNA and surrounding ions \cite{Kesselheim2014, Cui2010}. Understanding the origin of ionic current changes is critical for nanopore sensing techniques that discriminate and identify molecules on the basis of their ionic current signatures.

Controlled experiments on DNA-counterion interactions are enabled by designing molecules with DNA nanotechnology. A key asset of structural DNA nanotechnology is the ability to build structures with defined, programmable geometries at the nanometer scale. Using DNA origami techniques, a long, single-stranded scaffold can be folded into various three-dimensional shapes upon hybridization with short, custom-designed oligonucleotide staples \cite{Rothemund2006}. Synthetic nucleic acid structures with programmed mechanical and electrical properties can be used to probe mechanisms underlying transport in nanopores in a systematic manner \cite{Alibakhshi2017, Raveendran2018}. In this work, we use solid-state nanopores as a tool to study ion currents through DNA nanostructures consisting of bundles of four or sixteen dsDNA helices, complementing previous studies on ionic transport properties of DNA origami \cite{Plesa2014, Li2015}. Measurement of ion currents through multi-helix structures across a broad range of salt concentrations sheds light on the balance of mechanisms responsible for ionic current changes in nanopore translocation experiments.

\section{Methods}
Single-molecule translocation experiments were conducted using glass nanocapillary pores immersed in solutions of KCl as shown in Figure \ref{fgr:setup} and described in Supplementary Section S1. We characterized ionic transport through open nanocapillary pores by measuring the current-voltage (I-V) characteristics of nanopores in solutions of varying ionic concentration. The I-V data presented in Supplementary Figure S1 correspond to the same pores used to measure multi-helix bundle translocations in this study. The diameters of our conical pores can be estimated by applying a double-taper resistance model to the pore conductance extracted from the linear region of the current-voltage data \cite{Bell2015}. Fitting conductance data across all of the measured pores yields a mean pore diameter of \SI{14.2}{\nm}, which is larger than the diameters of the DNA nanostructures investigated in this work (Supplementary Figure S1b).

\begin{figure}[!p]
	\includegraphics[]{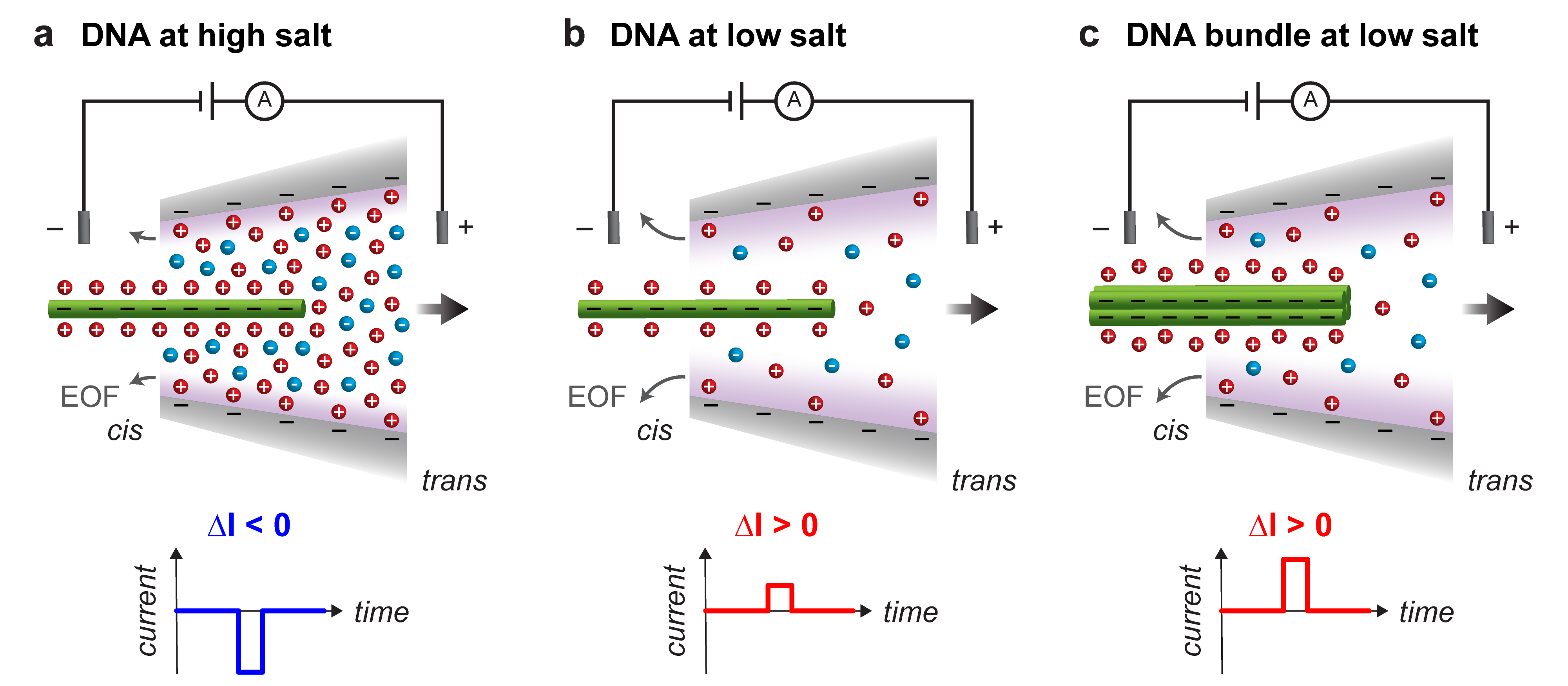}
	\caption{Nanopore sensing in high and low salt environments. DNA nanostructures are added to the reservoir containing the nanocapillary tip. Upon application of a voltage bias, the negatively-charged DNA nanostructures (green) translocate towards the base of the nanocapillary (in the direction of the black arrow). As each DNA nanostructure translocates through the pore, a transient ionic current modulation $\Delta I$ is observed. (a) At high salt concentrations typically used in nanopore sensors, counterions screen the glass surface charge within a region close to the pore (purple shading), generating small electroosmotic flow (EOF) out of the pore. Negative current changes, or blockades, are observed upon translocation of a single dsDNA strand. (b) At low salt concentrations, counterions screen the glass surface charge within a region further from the pore, leading to higher EOF. The introduction of mobile counterions by DNA outweighs frictional current reduction in the presence of a low background concentration of ions. Positive current changes, or enhancements, are observed upon translocation of a single dsDNA strand. (c) At low salt concentrations, the translocation of DNA bundles (composed of multiple dsDNA strands) leads to larger positive current changes compared to that of a single dsDNA strand.}
	\label{fgr:setup}
\end{figure}

\begin{figure}[!t]
	\includegraphics[]{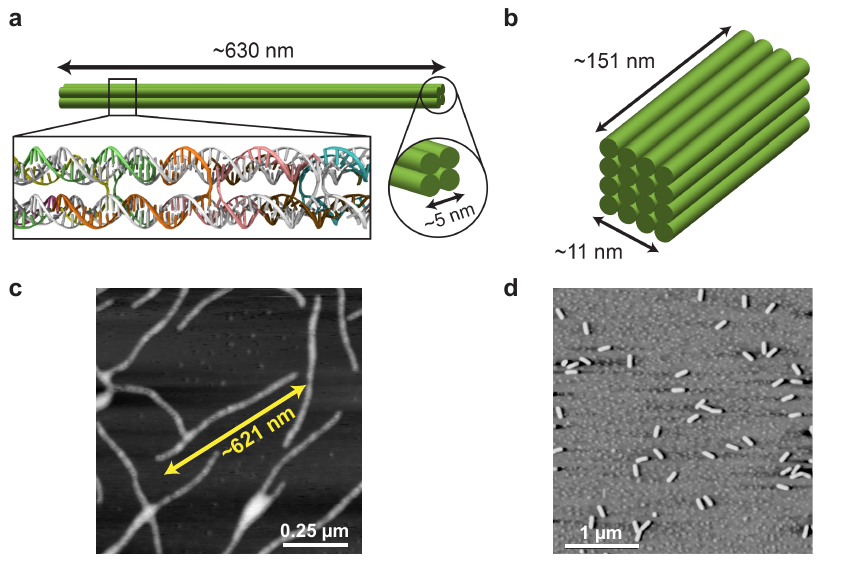}
	\caption{(a) Schematic representation of the 4-helix bundle (4HB) with approximate dimensions labeled on each side. The length is around \SI{630}{\nm} and the width is around \SI{5}{\nm}. The bottom left inset is the side cross-sectional view of an atomic-level illustration of a representative segment of the 4HB, as rendered in VMD using an all-atom structure generated from the caDNAno design \cite{Yoo2014, Humphrey1996}. The grey strand represents the scaffold. Colored segments represent different staple strands, portions of which cross over and connect adjacent parallel helices together. (b) Schematic representation of the 16-helix bundle (16HB), which has a length around \SI{151}{\nm} and a width around \SI{11}{\nm}. (c) Atomic force microscopy image of 4HB structures. (d) Atomic force microscopy image of 16HB structures.}
	\label{fgr:4HB}
\end{figure}

We designed and assembled DNA bundles composed of four (Figure \ref{fgr:4HB}a) or sixteen (Figure \ref{fgr:4HB}b) parallel dsDNA double-helices connected by periodically-repeating crossover strands. From hereon, we refer to these structures as 4-helix bundles (4HB) and 16-helix bundles (16HB), respectively. Synthesis procedures for the 4HB and 16HB are described in Supplementary Section S2, and sequences for the full sets of strands can be found in Supplementary Sections S9 and S10. Atomic force microscopy (AFM) imaging verified that both multi-helix DNA nanostructures folded successfully and according to design (Figures \ref{fgr:4HB}c and \ref{fgr:4HB}d). Further gel electrophoresis assays were performed to confirm stability of the structures across the entire range of salt conditions used in subsequent nanopore measurements. All gel electrophoresis results showed a clear monomer band corresponding to the correctly folded DNA nanostructure even after several hours of incubation at room temperature, indicating that DNA bundles remain intact in all of the buffer conditions considered in this work (Supplementary Section S3).

\section{Results}
\subsection{Capture rate enhancement with neutral polymers}
We studied the voltage-driven translocation of DNA bundles through nanopores by applying a constant voltage bias $V=\SI{600}{\mV}$ across the pore such that the negatively-charged DNA bundles translocated from the cis to trans reservoir (Figure \ref{fgr:setup}). It is important to note that measurement of DNA bundle translocations at low salt is complicated by the extremely low, if not near-zero, frequency of translocation. At a KCl concentration of \SI{0.25}{\Molar}, we were unable to observe any 4HB translocation events within half an hour of recording. It has been shown that electroosmotic flow (EOF) from glass nanopores hinders DNA entry into the pore, with the hindrance increasing at low salt concentrations \cite{Ermann2018a}. In glass capillaries, negative surface charge is responsible for outward EOF inside the pore accompanied by reversed EOF outside the pore \cite{Laohakunakorn2015}. As surface charge is also present in other types of pores, the presence of EOF is general to nanopores \cite{Waduge2017,Lu2013,Hoogerheide2014,Hoogerheide2009} and leads to effects like the reversal of electrophoretic transport depending on the zeta potentials of the translocating molecule and pore wall \cite{Firnkes2010, Reiner2010}. Dilute amounts of dynamically adsorbed uncharged polymers have been demonstrated to quench EOF in capillaries \cite{Hickey2009} and, more importantly, increase the frequency of single-molecule translocation events through nanocapillary pores in low salt conditions \cite{Ermann2018a}. In the latter study, EOF quenching in nanocapillaries was demonstrated with polyethylene glycol (PEG, MW = \SI[per-mode=symbol]{8000}{\gram\per\mole}), a neutral polymer commonly used for inert surface treatment of silica surfaces \cite{Alcantar2000}. We added PEG 8000 to the measurement solution on both sides of the nanopore to promote DNA bundle translocations in situations where the KCl concentration was too low to observe a sufficient number of translocation events over measurement times of several hours.

\begin{figure}[!h]
	\includegraphics[]{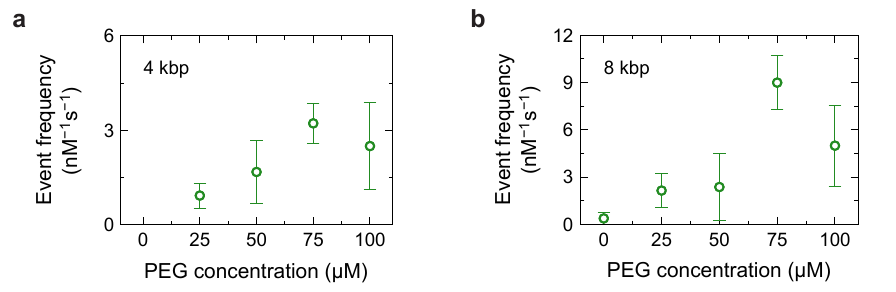}
	\caption{Adding PEG to the measurement solution increases the rate at which dsDNA translocates through a glass nanocapillary pore. Frequency of (a) 4 kbp dsDNA and (b) 8 kbp dsDNA translocation events with different concentrations of PEG 8000 added to the measurement solution. Each symbol and error bar represent the mean and standard deviation of event frequencies measured from several pores. In all experiments, an equimolar mixture of 4 kbp and 8 kbp dsDNA was measured. The detected events were identified afterwards as 4 kbp or 8 kbp translocations based on separation of the bimodal electronic charge deficit distribution, with 4 kbp dsDNA possessing a lower electronic charge deficit than 8 kbp dsDNA. Measurements were performed in \SI{0.35}{\Molar} KCl at \SI{600}{\mV}. Data were collected from $N$ = 4, 4, 5, and 8 pores for 25, 50, 75, and \SI{100}{\micro\Molar} PEG 8000 measurements, respectively. For reference, data on 8 kbp DNA translocation in KCl without PEG are shown in (b) at \SI{0}{\micro\Molar}.}
	\label{fgr:PEG}
\end{figure}

We determined the optimal concentration of PEG for enhancing the translocation frequency of DNA through a glass nanopore by measuring an equimolar sample of 4 kbp and 8 kbp dsDNA in KCl solutions with varying amounts of PEG 8000 added. We performed these measurements in \SI{0.35}{\Molar} KCl, the concentration at which the onset of EOF was empirically observed in glass nanopores of a similar size \cite{Ermann2018a}. As shown in Figure \ref{fgr:PEG}a, the translocation frequency of 4 kbp dsDNA increased with PEG concentration, although high translocation frequencies were more consistently observed in \SI{75}{\micro\Molar} PEG 8000. For 8 kbp dsDNA, \SI{75}{\micro\Molar} PEG 8000 also tended to yield the highest translocation frequencies on average, but pore-to-pore variability is a major contribution to the error bars in Figure \ref{fgr:PEG}b. When \SI{75}{\micro\Molar} PEG 8000 was added to the sensing medium, the translocation frequency of 8 kbp dsDNA reached \SI{9.00\pm 1.73}{\per\nano\Molar\per\second} (mean $\pm$ SD across five nanopores), which is over 20 times higher than the translocation frequency of \SI{0.37\pm 0.38}{\per\nano\Molar\per\second} observed in \SI{0.35}{\Molar} KCl solutions without PEG. Lower concentrations of PEG 8000 (\SI{25}{\micro\Molar} and \SI{50}{\micro\Molar}) do not appear to quench EOF as effectively since lower translocation frequencies were observed at these PEG concentrations. At high PEG concentrations, potential blockage of the pore by PEG molecules may hinder DNA translocation, as suggested by the lower mean translocation frequencies at \SI{100}{\micro\Molar} PEG 8000 than at \SI{75}{\micro\Molar} PEG 8000. Based on these results, we chose to add \SI{75}{\micro\Molar} PEG 8000 in subsequent measurements to promote DNA bundle translocations. Regardless, any concentration of PEG within the tested range can be used to enhance the capture of rate of DNA. The root-mean-square noise of the ionic current recording is almost constant with the addition of PEG, hence the detection of DNA translocations remains possible. Sample ionic current traces from nanopores in solutions with and without PEG are provided in Supplementary Section S4. Given our noise floor and choice of pore, the translocation of single PEG 8000 molecules cannot be resolved in the ionic current signal. The bimodal nature of the electronic charge deficit distributions further indicated that the detected events represented 4 kbp and 8 kbp dsDNA translocations. We note that the enhancement of capture rate by charge-neutral polymers is not specific to KCl or PEG 8000, as we have observed capture rate enhancement in solutions of LiCl \cite{Ermann2018a} as well as solutions of other charge-neutral polymers (Supplementary Section S5). However, we anticipate that the polymer size and concentration will have to be appropriately tuned to achieve notable capture rate enhancement in nanocapillary pores of different sizes.

\subsection{Translocation of bundled DNA nanostructures}
\begin{figure}[!t]
	\includegraphics[]{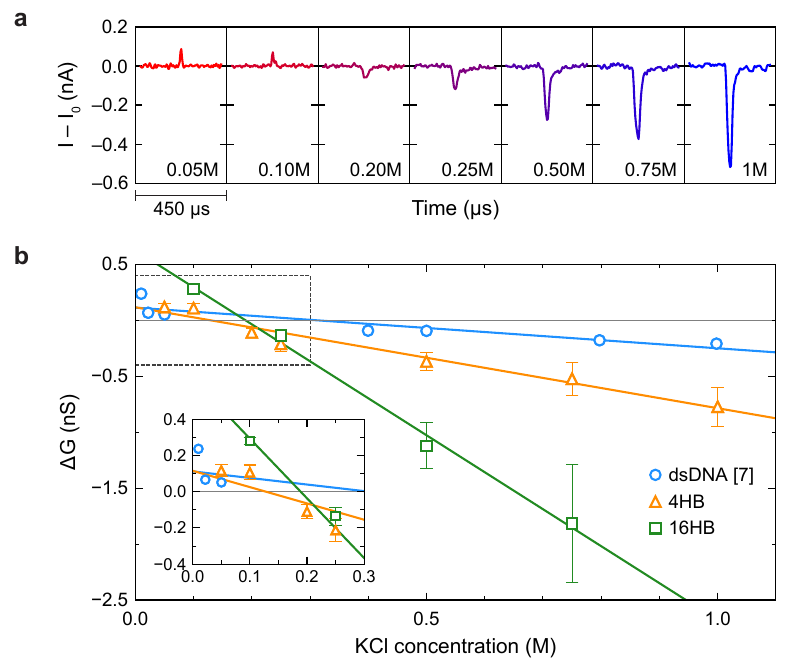}
	\caption{Salt-dependence of conductance changes observed for DNA bundle translocations at \SI{600}{\mV}. (a) Representative ionic current traces of 4HB translocations at several KCl concentrations with \SI{75}{\micro\Molar} PEG 8000 added to the measurement buffer. The baseline current was subtracted from each trace for better comparison. \SI{450}{\us} of the translocation event is shown for each trace. Enhancement of counterion concentration dominates at low salt  while frictional effects dominate at high salt. Correspondingly, current enhancements are observed at low salt while current blockades are observed at high salt. (b) Conductance change versus KCl concentration for dsDNA \cite{Steinbock2012a} (blue circles), 4HB (orange triangles), and 16HB (green squares) translocation. Linear fits yield crossover concentrations of approximately \SI{126}{\milli\Molar} (4HB, 95\% CI: [59,174] \SI{}{\milli\Molar}), \SI{187}{\milli\Molar} (16HB, 95\% CI: [147,211] \SI{}{\milli\Molar}), and \SI{310}{\milli\Molar} (dsDNA). Each symbol and error bar represent the mean and standard deviation of all events collected from multiple pores. Data were averaged from pores with and without PEG. The inset provides a more detailed view of the conductance data around the crossover points.}
	\label{fgr:dG}
\end{figure}

Demonstration that PEG can be used to increase the throughput of nanopore measurements at low salt enabled us to collect data on DNA bundle translocations across a broad range of salt concentrations. The salt-dependence of ionic current changes ($\Delta I$) due to the translocation of DNA bundles through nanopores is shown in Figure \ref{fgr:dG}. For both the 4HB and 16HB, we observed an inversion of the ionic current change similar to what has been observed for dsDNA. Namely, 4HB and 16HB translocations decrease ion currents at high salt concentrations but increase ion currents at low salt concentrations. Previous studies on dsDNA translocation through solid-state nanopores report crossover concentrations of \SI{0.34}{\Molar} and \SI{0.37}{\Molar}, with the conductance change varying approximately linearly with salt concentration \cite{Steinbock2012a,Smeets2006}. In glass nanocapillary pores, dsDNA was observed to produce current enhancements at KCl concentrations below \SI{0.3}{\Molar} \cite{Steinbock2012a}. For comparison, we show the conductance changes ($\Delta G = \Delta I/V$) due to dsDNA, 4HB, and 16HB translocations as a function of salt concentration in Figure \ref{fgr:dG}b. $\Delta G$ due to dsDNA translocation through glass nanocapillaries is extracted from literature \cite{Steinbock2012a}. Performing bootstrapped linear regression on the full set of translocation event data yields crossover concentrations around \SI{126}{\milli\Molar} (95\% confidence interval (CI): [59,174] \SI{}{\milli\Molar}) and \SI{187}{\milli\Molar} (95\% CI: [147,211] \SI{}{\milli\Molar}) for the 4HB and 16HB, respectively. For dsDNA, we extract a crossover concentration around \SI{310}{\milli\Molar} based on a least-squares linear fit of mean conductance change data reported for dsDNA translocation in the range of KCl concentrations shown in Figure \ref{fgr:dG} (\SI{0}{\Molar} to \SI{1}{\Molar} KCl) \cite{Steinbock2012a}. The crossover concentrations for both multi-helix bundles are lower than the crossover concentration for dsDNA, although the 16HB appears to have a higher crossover concentration than the 4HB. In other words, the crossover concentration decreases when the bundle size increases from one to four helices, but increases when the bundle size is further increased from four to sixteen helices. While there is some variance in the diameters of nanocapillary pores used to measure 4HB and 16HB translocations, pore size does not have a strong effect on the observed conductance changes (Supplementary Figure S7).

To investigate whether the use of PEG affects the measured conductance changes, we compared data on 4HB translocation in solutions with and without PEG across the full range of salt concentrations studied. At salt concentrations of \SI{0.25}{\Molar} and below, 4HB translocations were not observed in KCl solutions without PEG. Conductance change data in Supplementary Figure S8a show good agreement between values measured with and without PEG where possible, indicating that dilute PEG solutions can be used as a tool to reveal DNA translocations at low salt without affecting conductance changes. Although all of our measurements were performed at \SI{600}{\mV}, the trend in conductance change is also similar at different applied bias voltages (Supplementary Figure S8b). A final potential concern resides in the fact that low molecular weight oligomers are known to diffuse out of cross-linked PDMS \cite{Lee2003}. Since our fluid reservoirs are housed in millimeter-scale PDMS chambers, there is possible fluid contamination due to leaching of material from the PDMS cell. To exclude the possibility that uncrosslinked oligomers generated or influenced the translocation signals, we replicated our measurements in PDMS devices that had undergone an oligomer extraction treatment. Full description of this procedure can be found in Supplementary Section S1. We again find good agreement between conductance data obtained from the original and treated PDMS devices (Supplementary Figure S8c).

\section{Discussion}
For dsDNA, Smeets et al. originally explained the crossover in ionic current modulation with a simple model in which the conductance change is a combination of the decrease in ions excluded by the volume of the DNA and the increase in mobile counterions needed to neutralize charge on the DNA \cite{Smeets2006, Yoo2012a}. Subsequent calculations showed that the number of extra ions introduced by the surrounding counterion cloud exceeds the number of ions excluded from the pore by DNA in KCl solutions up to \SI{0.8}{\Molar} \cite{Kesselheim2014}. Considering changes in ion concentration in the sensing volume alone, the total ion current should increase during DNA translocation in KCl solutions up to this concentration. However, ion conductivity fundamentally depends on ion concentration and ion mobility. Molecular dynamics simulations of dsDNA translocation through nanopores have demonstrated that molecular friction between DNA and salt ions is predominantly responsible for current blockades by reducing ion mobility at the DNA surface \cite{Kesselheim2014}. Above the crossover concentration, this frictional current reduction outweighs direct current enhancement from the increase in counterions, thus causing the total ion current to decrease during DNA translocation. These two regimes are illustrated in Figure \ref{fgr:setup}: positive modulations are observed in the ionic current recording at low salt while negative modulations are observed in the ionic current recording at high salt.

The a priori expectation of the shift in crossover concentration with DNA bundle size is not obvious, as both the enhancement in counterion concentration and reduction in ion mobility due to friction at the DNA surfaces should increase concurrently with increasing bundle size. One might postulate that the predominant effect is the increase in concentration enhancement with bundle size. A high density of counterions in the porous regions between connected helices would increase the direct current contribution to the total current density, causing the crossover concentration to be higher. Our results on 4HB translocation refute this scenario and instead suggest that the frictional reduction of ionic current dominates the ionic current enhancement from increased counterion concentration. This downward shift in crossover concentration may occur if either ion mobility is lower near 4HB structures than it is near dsDNA, or the enhancement in counterion concentration at low salt is suppressed to a greater degree for 4HB structures than for dsDNA. The former scenario may occur if the movement of ions confined between grooves of adjacent DNA helices is impeded even more than that of ions around the backbone of a single dsDNA double-helix during translocation, thereby reducing the mean free path and effective mobility of ionic charge carriers. Additionally, DNA-ion interactions are dynamic processes in which counterions transiently associate and disassociate with the DNA \cite{Cui2011}. Inside the 4HB, fewer free counterions may be available to exchange with counterions adsorbed to the inner surfaces of the four helices, resulting in longer counterion residence times and slower migration. 

At low salt concentrations such as those below the crossover concentration, limitations imposed by the rate of ion diffusion into the DNA structure and its immediate vicinity should also be considered. From the steady-state solution of the Smoluchowski equation for a hemispherical sink, the diffusion-limited current is estimated to be \SI{119}{\pico\ampere} for the 4HB at \SI{0.05}{\Molar} KCl \cite{Hille2001}. Summing the experimentally measured current enhancement and estimated reduction in current due to the excluded volume of ions, the total current through the 4HB can further be estimated to be around \SI{127}{\pico\ampere} (more details regarding these approximate calculations can be found in Supplementary Section S8). Comparison of these current values suggests that the current through the 4HB approaches the diffusion-limited value at low salt concentrations, meaning that counterions cannot be supplied quickly enough to highly charged multi-helix bundle structures. Thus, the magnitude of current enhancement from multi-helix DNA structures is increasingly reduced as the ionic strength of the surrounding solution is decreased, which may also explain part of the reduction in crossover concentration.

Unexpectedly, the onset of negative current blockades occurs at a higher salt concentration for the 16HB compared to the 4HB. Given the decrease in crossover concentration for the 4HB relative to dsDNA, we would expect the crossover concentration to decrease yet again for the 16HB. One possible explanation for the discrepancy in trends is the differing aspect ratios of the 4HB and 16HB. The 4HB and 16HB were both folded from the same m13mp18 scaffold, meaning that the length of the 16HB is four times shorter than that of the 4HB. Unlike the 4HB, the end-to-end length of the 16HB (around \SI{151}{\nm}) is comparable to the effective sensing length of the nanopore \cite{Bell2016}. As a result, fewer DNA-ion interactions may be counted towards the total ionic current blockade, which could shift the crossover towards a higher concentration. When the molecule length is comparable to the sensing length, edge effects also begin to play a significant role, rendering direct comparison to experimental data on 16HB translocation with results on dsDNA and 4HB translocation difficult \cite{Meller2001a}.

The short dwell times of DNA bundles in nanopores raise questions regarding the ability to resolve ion behavior from DNA bundle translocation signals and interpret such signals in terms of time-averaged quantities like bulk concentrations and mobilities. The bulk ionic mobilities of potassium and chloride ions are similar with $\mu_\text{K} = \SI{7.62e-8}{\meter\squared\per\volt\per\second}$ and $\mu_\text{Cl} = \SI{7.91e-8}{\meter\squared\per\volt\per\second}$ \cite{Smeets2006}. Given an electric field around \SI{e7}{\volt\per\meter} at the tip of the nanocapillary pore \cite{Bell2016}, free potassium or chloride ions reach electrophoretic migration velocities up to \SI{762}{} or \SI{791}{\mm\per\second}. Based on the nominal bundle length and mean translocation time at \SI{600}{\mV}, the translocation velocity of the 4HB ranges from approximately \SI{4.10}{\mm\per\second} to \SI{28.5}{\mm\per\second} from \SI{1}{\Molar} to \SI{0.05}{\Molar} KCl. The translocation velocity of the 16HB ranges from approximately \SI{0.665}{\mm\per\second} to \SI{10.0}{\mm\per\second} from \SI{0.75}{\Molar} to \SI{0.1}{\Molar} KCl. Since the DNA bundles are relatively stationary compared to mobile ions even at the fastest translocation conditions, measurements of many DNA bundle translocation events can in fact capture information related to overall ion movements down the length of the bundle. In the future, models of DNA translocation through nanopores should be improved to account for the dependence of ionic current changes on arrangements of DNA structure in three dimensions. Molecular solvent structure and ion-specific short-range interaction potentials may affect the electrostatic and hydrodynamic behavior of three-dimensional DNA structures with closely-coupled charged strands \cite{Savelyev2007, Allahyarov2004, Yoo2013}, thus calling for more accurate models beyond those of standard continuum theories which may produce the present experimental results \cite{Bonthuis2013}.

\section{Conclusion}
In summary, we used nanopore sensors to study the salt-dependence of ion currents through bundled DNA nanostructures. We first demonstrated that neutral polymers such as PEG may be used to suppress electroosmotic outflow in order to promote capture and translocation of large, charged nanostructures through nanopores. This technique enables faster collection of single-molecule translocation data and recording of events that would otherwise take a long time to observe. We measured DNA bundle translocations across a wide range of salt concentrations and found that ionic current modulations invert at a lower crossover concentration for four- and sixteen-helix DNA bundles than for a single dsDNA strand. Remarkably, the reduction in crossover concentration is greater for the four-helix bundle than for the sixteen-helix bundle. These trends suggest that ion mobility is reduced in the presence of bundled DNA structures when compared to single dsDNA helices. Moreover, ionic current modulations depend on both the shape and size of translocating nanostructures, as evidenced by the opposing trends in crossover concentration for DNA bundles of increasing diameter but decreasing length. Our observations on ion movements through DNA bundles yield further insight into the mechanisms responsible for the electrical signatures of molecules translocating through nanopores. These results may be relevant to the design of hybrid DNA origami nanopore sensors \cite{Plesa2014} or DNA carriers for biosensing applications \cite{Bell2015} in which a high signal-to-noise ratio in a certain ionic environment is desirable. Finally, we believe experimental measurements of rationally-designed, model DNA nanostructures may be used as a reference to calibrate or validate molecular dynamics simulations and atomic-level models of DNA under electric fields \cite{Aksimentiev2010}. Such contributions not only inform the development of DNA origami-based nanodevices that aim to control ion flux (e.g. synthetic membrane channels \cite{Langecker2012}), but also improve the fundamental understanding of ion transport in confined environments.

\begin{acknowledgement}
The authors thank Jinglin Kong for providing the AFM image and materials for the 16-helix DNA bundle, as well as Christian Holm and Kai Szuttor for useful discussions. V. W. acknowledges support from the Winston Churchill Foundation of the United States. N. E. acknowledges funding from the EPSRC, Cambridge Trust and Trinity Hall, Cambridge. U. F. K. acknowledges funding from an ERC Consolidator Grant (DesignerPores No. 647144).
\end{acknowledgement}

\begin{suppinfo}
Nanopore fabrication and measurement; DNA nanostructure synthesis; gel electrophoresis characterization of DNA nanostructures; sample ionic current traces; capture rate enhancement with other salts and molecules; conductance data from different PDMS devices and pores; conductance data at different voltages; staple strand sequences for DNA nanostructures (pdf).
\end{suppinfo}

\bibliography{Paper_DNABundles}

\end{document}